# Computational and experimental characterization of NF023, a candidate anticancer compound inhibiting cIAP2/TRAF2 assembly

Federica Cossu[1,2,*], Luca Sorrentino[1,2,*], Elisa Fagnani[1], Mattia Zaffaroni[2], Mario Milani[1,2], Toni Giorgino[1,2,+], and Eloise Mastrangelo[1,2,+]

[1] Istituto di Biofisica, Consiglio Nazionale delle Ricerche (CNR-IBF), Via Celoria, 26, I-20133 Milan, Italy

[2] Dipartimento di Bioscienze, Università degli Studi di Milano, Via Celoria, 26, I-20133 Milan, Italy

* These authors contributed equally to the work

+ Co-corresponding authors. E-mail: toni.giorgino@cnr.it, eloise.mastrangelo@unimi.it

**Abbreviations:** IAPs (Inhibitor of Apoptosis Proteins); cIAPs (cellular IAPs); XIAP (X-linked IAP); BIR (Baculovirus Iap Repeat); TRAF2 (TNF receptor-associated factor 2); MST (Micro-Scale Thermophoresis); MD (Molecular Dynamics); NF-κB (Nuclear Factor kappa-light-chain-enhancer of activated B cells).

# Abstract

Protein-protein interactions are the basis of many important physiological processes, and are currently promising, yet difficult, targets for drug discovery. In this context, inhibitors of apoptosis (IAPs)-mediated interactions are pivotal for cancer cell survival; the interaction of the BIR1 domain of cIAP2 with TRAF2 was shown to lead the recruitment of cIAPs to the TNF-receptor, promoting the activation of NF-κB survival pathway. In this work, using a combined *in silico-in vitro* approach we identified a drug-like molecule, NF023, able to disrupt cIAP2 interaction with TRAF2. We demonstrated *in vitro* its ability to interfere with the assembly of the cIAP2-BIR1/TRAF2 complex and performed a thorough characterization of the compound's mode of action through 248 parallel unbiased molecular dynamics simulations of 300 ns (totaling almost 75 μs of all-atom sampling), which identified multiple binding modes to the BIR1 domain of cIAP2 *via* clustering and ensemble docking. NF023 is, thus, a promising protein-protein interaction disruptor, representing a starting point to develop modulators of NF-κB-mediated cell survival in cancer. This study represents a model-procedure that shows the use of large-scale molecular dynamics methods to typify promiscuous interactors.

# Introduction

Over the past few years, modulation of protein-protein interactions (PPI) is becoming a crucial target in the drug discovery process. PPI inhibitors can be designed to achieve highly specific binding to interfere with the formation of a target protein complex [1]. Inhibitor of Apoptosis Proteins (IAPs) are a family of proteins whose overexpression correlates with tumor onset, progression, invasion, chemoresistance and poor prognosis [2]. IAPs-mediated protein-protein interactions are pivotal for cancer cells survival and represent optimal targets for cancer therapy. IAPs contain one to three BIRs (Baculovirus IAP Repeat) domains [3] which can be classified into type I (namely BIR1) and type II (BIR2 and BIR3), based on the absence or presence of a distinctive cleft, called IBM (IAP Binding Motif) groove [3]. As a consequence, type I and II BIRs use different regions of their surface to establish specific interactions with a variety of cellular partners to inhibit apoptosis and promote cell survival. Anti-cancer drug discovery studies have mostly targeted type II BIRs with 'SMAC-mimetic' inhibitors but unfortunately mechanisms of resistance have been detected in some cancer cell lines [4].

Cellular IAPs (cIAP1 and cIAP2) are involved in the tight regulation of the canonical and non-canonical NF-κB (Nuclear Factor kappa-light-chain-enhancer of activated B cells) pathways [5]. In this context, cIAPs act as E3 ligases ubiquitinating RIPK1 (Receptor-Interacting serine/threonine-Protein Kinase 1) and NIK (NF-kappa-B-Inducing Kinase) promoting cell survival and preventing the release of pro-inflammatory molecules, respectively. Type I BIRs mediate cIAPs' recruitment to the TNFR1/2 (tumor necrosis factor receptors 1 and 2) [6] *via* the interaction with TRAF2 (TNF receptor-associated factor 2, UniProt ID Q12933), being such interaction strictly required for cIAPs' E3 ligase activity [7,8]. The homologous BIR1 domain of XIAP (X chromosome-linked IAP), upon homo-dimerization, was shown as pivotal for its association with TAB1 (TGF-Beta Activated Kinase 1 Binding Protein 1), and the downstream activation of NF-κB [8]. The molecular and structural details of IAPs BIR1-mediated interactions have been exhaustively described [7–10]. The mechanisms of resistance to type II BIRs-directed therapies depend on the renewed cIAP2 E3 activity and its positive feedback regulation [4]. Thus, BIR1-mediated protein-protein interactions represent an innovative target to find alternative or supporting strategies to induce cell death.

In this work, we tackled the assembly of type I BIRs-mediated complexes to find NF-κB modulators through a structure-based computational screening which yielded NF023, a suramin analog, as a candidate for targeted optimization (Figure 1A). We describe the thorough functional and structural characterization of NF023 antagonizing cIAP2-BIR1:TRAF2 complex formation, conducted through the integration of docking and hundred-microsecond-scale, explicit-solvent molecular dynamics simulations. We used an ensemble of 248 unbiased, free ligand diffusion trajectories to generate hypotheses on binding modes, and assess the heterogeneity of the ensemble statistically with respect to preferred contact residues and over-represented poses. We finally mapped the compound's binding mode *in vitro* through point mutations and Microscale Thermophoresis (MST) studies, with the aim of providing information for the structure-based design of novel PPI disrupting molecules specific for BIR1 domain as pro-apoptotic agents.

# Results

## Identification of NF023 as a cIAP2-BIR1 ligand

NF023 (Figure 1A) is an analog of the drug suramin, currently the treatment of choice against parasites causing African sleeping sickness (African trypanosomiasis) and river blindness (onchocerciasis) [11,12].

The use of NF023 as a model for TRAF2:cIAP2 protein-protein assembly antagonism was supported by the fact that NF023 is a known ligand of XIAP-BIR1, a domain analogous in structure and function to cIAP2-BIR1 (although with a different multimerization state)[13]. Additionally, a preliminary docking screen of the LOPAC library targeting only the interfacial region of cIAP2-BIR1 engaged in the pro-survival PPI with TRAF2 yielded NF023 among the 6 top scoring hits (Figure 1B, Supplementary Figure S3 and Supplementary Text S1). In the best-scoring conformation (Figure 1C, in cyan) NF023 is found interacting with the α1-α2 region of cIAP2-BIR1, running parallel to the TRAF2 axis in the complex. The second-best pose (Figure 1C, in light blue) comprises bended NF023 conformations, virtually able to intercept α1 and partially α2, but also to reach the β-sheet region. These preliminary data prompted a thorough biophysical characterization of the mode of action of NF023 through a combined *in vitro* and *in silico* approach.

## Free-binding MD trajectories

NF023 is a heavily sulfonated compound for which promiscuity of binding is expected (Supplementary Text S2) [14]. In order to derive a detailed picture of NF023 binding to cIAP2-BIR1 and characterize hotspots exploitable in drug development we performed and analyzed an extensive set of 248 unbiased, all-atom, MD simulations, totaling approximately 75 μs of sampling (Figure 2A).

First, we considered the final frames ($t$ = 300 ns) of each of the 248 replicas (see Materials and Methods), computing how frequently each chemical group of NF023 was within 4 Å of a residue of BIR1 (definition of a contact). The first feature of note is the presence of two diffuse contact surfaces, namely in the α1-α2 region (residues M25-A53) and around the β3-α3 loop (residues K77-K85) (Figure 2B). The α1-α2 region is a compact bundle of two α-helices directly facing TRAF2; the β3-α3 loop likely constitutes a hot-spot because of its relatively large solvent exposed-surface, entropically favored by the loop flexibility. For both surfaces the protein-ligand binding is mainly driven by hydrophobic interactions, established by NF023's non-polar groups and aromatic moieties (Figure 2C).

Secondly, significant contributions to the protein-ligand polar interactions are provided by NF023's negative moieties (the sulfonic groups and carbonyl groups), which are frequently accommodated in the cIAP2-BIR1 region lined by residues of α2 helix S46, R48 and R52, previously identified as TRAF2 interface residues [7]. The amphiprotic region S46-E47-R48-R52 may therefore be a reasonable guess as the location of the drug's center of symmetry, characterized by an amine-carbonyl-amine arrangement. The hypothesis is supported by the distribution of contacts observed for the central group in Figure 2C. Secondary or transient binding sites might exist around the already mentioned β3-α3 loop, although lacking a clear signature, and possible artifacts due to structure truncation at the C-terminus.

## Final-frame ensemble clustering

The availability of an ensemble of trajectories allowed the use of statistical methods, described in the Methods section, to uncover a complex binding landscape consistent with the highly charged and flexible NF023 molecule. The clustering procedure on the ensemble of final conformations with a cut-off of 3 Å RMSD in ligand poses identified four distinct binding clusters, here labelled as C1 to C4 (Figure 3). Significantly, two of the four (cluster C2 and cluster C3) are at the interface between BIR1 domain and TRAF2. In these clusters the compound is located between the α2 and α3 helices, at the bottom side (facing the C terminal ends of TRAF2) of the interface surface. Clusters C1 and C4 are farther from the TRAF2 interface; in both clusters the compound is located in proximity of the C-terminus, inside a long cleft formed around α2 and β1.

# Ensemble redocking

The ensemble re-docking procedure, designed to uncover possible induced fit and cryptic pockets, uncovered three binding poses with a predicted overall free energy value as low as -9.5 kcal/mol. The corresponding poses, shown in Figure 4, are labelled as R1, R2 and R3. Of the three, in both the R2 and R3 poses NF023 clearly overlaps with the interface between BIR1 and TRAF2, in configurations similar to those previously found in C2 and C3. R1 is also near the interface with the partner protein, slightly shifted towards the C-terminus, as found in C1. In all poses (available as Supplementary Data) the ligand is accommodated by clefts on the cIAP2 surface.

# NF023 has multiple interaction regions on BIR1

The results of the free-binding simulations, taken together with the final-frame clusters support the promiscuous nature of NF023 binding to the protein surface. Starting both from the considerations on the protein residues that more frequently form contacts with NF023 and from the clusters analysis, we identify the main regions on the BIR1 domain surface involved in ligand interactions. In particular, we propose the partial decomposition of the cIAP2-BIR1 surface in the three extended regions, namely "T" for the interface with TRAF2; "U" ("upper" region) for the patch facing towards the N termini of TRAF2 coiled-coil helices; and "R" ("reverse") for the region roughly coinciding with BIR1's C-terminal (Figure 5 and Table S2). The regions are overrepresented in the contacts distribution analysis. Two of the four clusters identified by the ensemble clustering procedure (clusters 2 and 3) are at the "T" region, thus likely disruptors of the protein-protein interaction, while clusters 1 and 4 lie respectively on the "R" and "U" regions. Analogously, two of the three poses with higher affinity predicted by the ensemble redocking procedure lie in the "T" region.

# Mapping NF023 binding to cIAP2-BIR1 *in vitro*

In the light of the results obtained from the *in silico* docking and from the MD simulation, and to validate the binding modes of NF023, we performed mutagenesis experiments on cIAP2-BIR1. In particular, we inserted point mutations in the α1 region (positions 28, C>S; 31, Y>A; 34, S>A), in the α2 region (48, R>A) and in the β1 strand (56, Y>A). The β3-α3 loop region is highly mobile (as shown by MD experiment) and was thus discarded from protein point-mutation selection. The choice of point mutations was also guided by available data reported in Zheng et al. [7], where cIAP2-BIR1 mutants were assessed for their ability to retain TRAF2 during pull down experiments.

The mutant variants of cIAP2-BIR1 displayed biophysical properties similar to the wild type protein in terms of hydration shape and thermal stability (Table S1 and Supplementary Figures S6 and S7), except for R48A, which displayed the lowest melting temperature value (12.7 °C lower than the wild type), indicating poor stability. NF023 affinity for wild type and mutant forms of cIAP2-BIR1 was measured using MST (see Supplementary Figure S5 and Material and Methods). In this experimental context, the most unstable mutant, R48A, did not show a suitable fluorescence signal and could not be tested, probably due to the buffer conditions of the labeling procedure. The $K_d$ values and standard deviations were calculated over three independent experiments. As reported in Table 1, NF023 binding affinity for the Y31A mutant was comparable with that for the wild type protein, whereas the C28S mutation led to a 100-fold decreased affinity. S34A and Y56A substitutions prevented NF023 binding, indicating that the side chains of such residues are critical for cIAP2-BIR1/NF023 adduct formation.

These observations are consistent with the MD results, which identified the α1-α2 region of cIAP2-BIR1 (residues M25-A53) as one of NF023 targeting hotspots. Furthermore, these results suggest that S34 (α1) and Y56 (β1) are critical target residues of cIAP2-BIR1 for the binding of NF023.

# NF023 impairs cIAP2-BIR1 binding to TRAF2

Both crystallographic and ITC experiments reported on the cIAP2-BIR1:TRAF2 complex showed a ratio of 1:3 [7]. When subjected to SEC, the TRAF2 construct displayed a non-homogenous peak suggesting the presence of different oligomeric states. The peak corresponding to the trimeric form was isolated for MST experiments with cIAP2-BIR1 and its mutant variants. Resulting $K_d$ values are reported in Table 1. Wild type cIAP2-BIR1 displayed a dissociation constant for TRAF2 of 39 ± 17 μM (Supplementary Figure S5). Comparable values were observed for the C28S and Y56A mutants.

The C28 residue is located close to the N-terminal of cIAP2-BIR1 construct (Met 25 being the first residue). We can hypothesize that this part of the protein - alone or in presence of the C28S mutation - can adopt different conformations with respect to that observed in the crystal structure, allowing the interaction of residue 28 with TRAF2. Analogously, residues S27 and E29, even though not in direct contact with the partner protein, when modified to Ala showed an enhancement of the affinity for TRAF in pull down experiments [7]. The substitutions Y31A and S34A resulted in the loss of TRAF2 binding, also confirming the contacts observed in the crystal structure (PDB ID: 3M0A) and the results previously obtained with pull down assays [7].

Excluding the S34A mutant, the other variants were incubated with NF023 at a concentration of 500 μM (5000-fold excess over the protein) prior to MST measurements with TRAF2, in order to assess the affinity of the protein–protein complex upon NF023 addition (Table 1). The presence of NF023 prevented wild type cIAP2-BIR1 binding to TRAF2 (as the thermophoresis of labelled cIAP2-BIR1 did not change upon increasing concentrations of TRAF2, when the labelled target was pre-incubated with 500 mM of NF023, Supplementary Figure S5, C), whereas the C28S and Y56A mutants, which displayed a much reduced affinity for NF023, retained their ability to engage TRAF2. As expected, no binding with TRAF2 was measured in the case of Y31A mutant, also in the presence of NF023.

# Discussion and conclusion

IAPs are excellent targets for cancer therapy and several therapies with SMAC-mimetics (SM) compounds are in advanced phases of clinical trials [15]. Such small molecules target type II BIR domains, thus preventing the pro-survival activity of IAPs. Nevertheless, some cases of overcoming resistance to SM treatment have been described [4,16], mainly involving upregulation of cIAP2. Since the pro-survival activity of cIAP2 depends on BIR1 mediated interaction with TRAF2, we conducted a characterization of the hot spots on the BIR1 domain through a combined *in silico* and *in vitro* approach leveraging a high-throughput MD setup [17,18]. Previous data on homologous domains and a docking screen prompted the use of NF023, an analog of suramin (Supplementary Text S2), to probe its interaction profile with cIAP2 and ideally inform future drug development efforts. The statistical analysis of cIAP2-BIR1/NF023 recognition by unbiased MD simulations and ensemble docking allowed the identification of three main hot spots: (*i*) the α1-α2 region, (*ii*) the β1 strand (residue Y56), and (*iii*) the β3-α3 loop. *In vitro* experiments confirmed that the α1-α2 region is crucial for the binding of both NF023 and TRAF2, being S34 an essential residue at the base of such interactions. For this reason, S34 represents a hotspot for the rational design of NF023 derivatives or other PPI inhibitors. The region of the β1 strand can also be an anchoring point for the design of cIAP2-BIR1 ligands, due to the presence of Y56 charged and exposed moiety, within a well-defined secondary structure element.

The behavior of the C28S mutant (mutation located in the "U" region, at the beginning of α1) is especially intriguing: it retains the ability to bind both TRAF2 and (to a reduced extent) NF023, yet pre-incubation with the inhibitor does not hamper the formation of a cIAP2-C28S:TRAF2 complex. The C28S substitution thus confers resistance to protein-protein inhibition by NF023. A possible hypothesis may be set forth in the light of MD results: the mutation leads to the selection of one of the alternative NF023 binding modes that does not compete with TRAF2.

Taken together, these results provide hints for the identification of hotspots on BIR domains and of chemical moieties for the design of further BIR1-directed candidates accessing the PPI site, as well as demonstrating the power of large-scale *in silico* methods for the characterization of large protein-protein inhibitors with complex binding patterns.

# Experimental section

## Preliminary docking screen

A preliminary virtual screen with the Library of Pharmacologically Active Compounds (LOPAC; 1280 commercially available compounds; www.sigmaaldrich.com) on cIAP2-BIR1 (PDB ID: 3M0D) produced a ranked list of 1249 compounds, with predicted binding free energy values (ΔG) ranging between -0.66 kcal/mol and -9.27 kcal/mol (Supplementary Figure S3 and Supplementary Text S1). The missing compounds showed positive ΔG. The screen used AutoDock4 [19] with a docking grid of 28.5 × 28.5 × 22.5 Å$^3$, centered on Met33's side chain, set to encompass the cIAP2-BIR1 surface involved in the interaction with TRAFs [7].

## Molecular dynamics

The coordinates of the cIAP2-BIR1 domain were taken from chain D of the crystal structure of the TRAF2:cIAP2 complex deposited with PDB ID 3M0A [7] (residues Met25 to Ser98). Titration states of side-chains were assigned at pH 7 via the ProteinPrepare algorithm [20], and the chain was capped with neutral acetylated and N-methyl termini. Protonation states for the zinc-coordinating residues were adjusted according to its metallic center geometry (i.e. His86 protonated at Nδ). The protein was parameterized with the Amber ff14SB force-field [21]. For NF023, each of the 6 $SO_3$ groups was taken to be negatively charged, according to the expected titration state at pH 7. Partial charges were assigned with the RESP method of the Antechamber package, and the GAFF2 force-field was used for parameterization [22,23]. The Marvin suite (v. 19.2) was used for drawing and characterizing chemical structures (www.chemaxon.com).

## Preparation of initial configurations

In order to optimize the *in silico* exploration of the configuration space of bound and unbound configurations, we prepared five systems (*generators*), each containing one cIAP2-BIR1 domain plus one NF023 molecule in the solvent at a different location and orientation. Initial coordinates for NF023 were extracted from PDB ligand ID 0BU. The five systems were produced by translating the ligand's center at a randomly-selected point on a 34 Å-radius sphere centered in BIR1 (Figure 2A). Each system was solvated with TIP3P water in a 73 × 73 × 73 Å³ box, neutralized and ionized to 150 mM NaCl ionic strength, and energy-minimized to reduce steric clashes. The systems were pre-equilibrated in constant pressure and temperature conditions (1 atm, 300 K) for 1 ns, with harmonic restraints for protein's and drug's heavy atoms linearly tapered towards zero [24].

System preparation and building were performed with the HTMD package [25]. All of the simulations were run with the ACEMD software [26] on the GPUGRID distributed computing network [17] with a 4 fs integration time-step enabled by the hydrogen mass repartitioning scheme [27].

## Large scale MD of cIAP2-BIR1/NF023

The equilibrated configuration of each generator was used as the initial condition of 50 replicas, i.e. 250 production runs of 300 ns each in the constant-volume ensemble with the Langevin thermostat. During each run both the protein and the ligand were unrestrained and, as such, they exhibit configuration changes as well as translational and rotational diffusion in the simulation box. The simulations comprise a grand total of 248 × 300 ns ≃ 75 μs of unbiased sampling (discounting two replicas not returned by the volunteer-based distributed system). Of note, replicas spawned from each generator start from the same initial configuration but their trajectories quickly decorrelate because of the stochasticity of the Langevin thermostat.

We restricted the structural analysis to the final frames of each trajectory for the, in order to maximize the independence of the samples of the "bound" configuration space thus realized. The diffusion of NF023 on the protein surface is relatively slow with respect to the sampled time scales of each trajectory (see e.g. Shan et al. [28] for millisecond-MD perspective, albeit for a different drug class). Once made contact with the surface, which happens within a hundred of ns (Supplementary Figure S1), the ligand's diffusion is slowed down, and it does not visit many more states in the simulation-accessible timescales (Supplementary Figure S2). Hence, frames of the same trajectory after initial contact are pretty much correlated, and it makes sense to only pick one; picking more frames would introduce a statistical bias in the analysis: namely, one would oversample the states visited by replicas which, by chance, made earlier contact. In other words, given the accessible timescales and availability of data, (approximate) ensemble averaging is preferable to time average. Note that trajectories spawned by the same initial conditions decorrelate quickly as a consequence of the randomization provided by the thermostat.

## Clustering of the ensemble of final MD configurations

In order to determine the poses of NF023 most statistically represented at the end of each of the 248 trajectories, we searched for clusters of compound orientations. More specifically, a matrix containing the RMSD values calculated from the pairwise structural comparison of the final frames was built, accounting for symmetry. A pairwise maximum RMSD of 2.0 Å and minimum population of 3 replicas per cluster was used as a criterion for the selection of replicas belonging to the same cluster. The coordinates of the sulphur atoms and the carbons of the central group of NF023 were used for the RMSD calculation, accounting for the chemical symmetry of the compound.

## Ensemble re-docking analysis

To address the high flexibility of NF023 (12 rotatable bonds) and the presence of cryptic pockets, possible induced-fit or conformational selection with intermediate, transient states of cIAP2, we employed a MD-based re-docking strategy (Figure 4) [29,30]. In particular, we extracted the final configurations of each MD replica, removing the ligand and explicit water, and used the resulting structure as the target for AutoDock Vina-based docking of NF023 [31] ; the focus grid was set to encompass the whole domain surface. The redocking analysis thus takes better account of the variability of the protein surface dynamics (induced fit effect), as different cavities form and disappear over time. For each configuration we selected the three best-scoring structures reported by the docking procedures, as well as their predicted affinity.

## Cloning and mutagenesis

The cDNA coding for human cIAP2 was retro-transcribed from a pool of human mRNAs. The genomic sequence of cIAP2-BIR1 (coding for amino acids 26-102) was cloned in pET-28b vector (Novagen) between NdeI and XhoI restriction sites. The clones for cIAP2-BIR1 point mutants were designed and produced with Q5® Site-directed mutagenesis kit (New England Biolabs). Primers used for mutagenesis experiments were as follows:

| C28S | Fwd | 5' – TATGTTGTCAAGTGAACTGTACC… |
|------|-----|-------------------------------|

|  |  | ...GAATGTCTACGTATTCCACTTTTCCTGCTGG – 3' |
|  | Rev | 5' – TGGCTGCCGCGCGGCACC – 3' |
| Y31A | Fwd | 5' – ATGTGAACTGGCGCGAATGTCTACGTATTCCAC – 3' |
|  | Rev | 5' – GACAACATATGGCTGCCG – 3' |
| S34A | Fwd | 5' – GTACCGAATGGCGACGTATTCCAC – 3' |
|  | Rev | 5' – AGTTCACATGACAACATATGG – 3' |
| R48A | Fwd | 5' – TGTCTCAGAAGCGAGTCTTGCTCGTGC – 3' |
|  | Rev | 5' – GGAACCCCAGCAGGAAAA – 3' |
| Y56A | Fwd | 5' – TGCTGGTTTCGCCTACACTGGTGTGAATG – 3' |
|  | Rev | 5' – CGAGCAAGACTCCTTTCTG – 3' |

## Chemicals

NF023 (8,8′-[Carbonylbis(imino-3,1-phenylene carbonylimino)]bis(1,3,5-naphthalene-trisulfonic acid) hexasodium salt hydrate) was purchased from Sigma–Aldrich, dissolved at 100 mM in ddH$_2$O and stored at -20 °C.

## Expression and purification of cIAP2-BIR1 forms and TRAF2

Plasmids for cIAP2-BIR1 expression were used to transform *Escherichia coli* BL21-CodonPlus (DE3)-RP competent cells (Agilent®). Protein expression was induced by adding isopropyl-β-D-thiogalactopyranoside (IPTG, Sigma-Aldrich) to a final concentration of 0.5 mM. After induction, bacteria were grown in LB medium additioned with 50 mg/L kanamycin for 3 h at 37 °C. Cells were harvested, resuspended in lysis buffer containing 50 mM Tris-HCl pH 7.5, 200 mM NaCl and protease inhibitors, treated with 20 μg/ml DNAse (Sigma-Aldrich), 40 mM MgSO$_4$, 100 μg/ml lysozyme for 30 minutes in ice and then disrupted by sonication. After elimination of debris by high speed centrifugation,

cIAP2-BIR1 forms were purified using Ni-NTA (HisTrap™ FFcrude, GE Healthcare), using 250 mM imidazole for their elution. After a gel filtration step (Superdex 75, GE Healthcare), proteins were stored in 50 mM BIS-TRIS pH 7.4, 200 mM NaCl, 5% glycerol and 10 mM Dithiothreitol (DTT). Concentration of protein fractions was performed with Amicon Ultra centrifugal filters (3 kDa cut-off).

The vector for the expression of TRAF2 construct (coding for amino acids 266-329) was gently provided by Prof. Hao Wu and expressed and purified as already reported [7]. Briefly, the crude extract was clarified and purified by affinity chromatography using a TALON column (GE Healthcare). The final SEC step was performed using a Superdex 75 10/300 Increase (GE Healthcare). Since this TRAF2 construct does not contain any tryptophan or tyrosine residues, quantification of protein samples was performed by their absorbance at 205 nm (calculating the extinction coefficient on the base of the sequence [32], 261790 $M^{-1}$ $cm^{-1}$). SEC fractions were kept separated to assess their purity by SDS-PAGE and their homogeneity by Dynamic Light Scattering. Fractions eluting with an apparent MW corresponding to the trimeric assembly were concentrated and used for Microscale Thermophoresis (MST) experiments.

## Dynamic light scattering

Purified proteins were centrifuged at 13,000 g for 10 minutes at 4 °C prior to DLS analysis. All measurements were performed in triplicates at 10 °C in a DynaPro instrument (Protein Solutions, CA, USA) with a data acquisition time of 30 sec. 20-25 single measurements were used for data averaging. The acquired data were analyzed using the software DYNAMICS V6 (Wyatt Technology) and the molecular weight corresponding to the obtained hydrodynamic radius was calculated with the software DYNAMICS V5 (Wyatt Technology). DLS data showed that all the proteins were less than 20% polydisperse [33], displaying the hydrodynamic radii reported in Table S1.

# Size exclusion chromatography

Analytical size exclusion chromatography (SEC) experiments were performed at different cIAP2-BIR1 concentrations (0.09 and 1.60 mM) in order to characterize wild type and mutant forms. During the SEC experiments sample volumes of 50 µl were injected on a Superdex 75 10/300 Increase column (GE Healthcare) attached to an ÄKTA Pure system in a buffer containing 50 mM BIS-TRIS pH 7.4, 200 mM NaCl and 10 mM DTT. Low molecular weight standards (GE Healthcare) were used to calibrate the column. Elution volumes ($V_E$) together with calculated molecular weights are reported in Table S1.

# Thermal Stability Assay

Thermal denaturation assays were conducted in a MiniOpticon Real Time PCR Detection System (Bio-Rad), as already reported [34]. The final protein concentration was 40 µM (60 µM in the case of TRAF2); sample plates were heated from 20 to 95° C (0.2 °C/min). Fluorescence intensity was measured within the excitation/emission ranges 470-505/540-700 nm (Table S1).

# Microscale Thermophoresis (MST)

MST experiments were performed with a Monolith NT.115 instrument, following the procedures recommended by the manufacturer (NanoTemper Technologies), as already reported [35]. Briefly, purified cIAP2-BIR1 forms were covalently labelled on Lysine residues with the Protein Labelling Kit RED-NHS 2nd generation (NanoTemper Technologies) and then used at the concentration of 100 nM. Dilution series were prepared to have the following concentration ranges: 300 nM-10 mM for NF023 [35]; 4 nM-133 µM for the TRAF2 protein *vs* wild type cIAP2-BIR1; 1.2 µM-40 mM for NF023 *vs* cIAP2-BIR1 mutants. Assays were performed at 24 °C in 50 mM Tris-HCl pH 8.0 containing 200 mM NaCl, 10% glycerol, 10 mM DTT and 0.05% Tween-20, using premium treated glass capillaries and setting 20% LED intensity and medium MST power.

# Author contributions

FC and LS contributed equally to this work. FC, LS, EF and MZ performed plasmid construction, tests for the *in vivo* production studies, and biophysical characterization. FC and MM performed the *in silico* virtual screening experiments. EF and TG performed the molecular dynamics simulations. TG and EM jointly coordinated the analysis. All authors approved the final manuscript.

# Acknowledgements

EM, EF and FC are thankful to AIRC foundation for the financial support (Grant no. MFAG17083), to Prof. H. Wu (Boston Children's Hospital, Boston, MA 02115, USA) for providing the vectors for the TRAF2 segment, to Dr. D. Tarantino for support in MST measurements and analysis, to Francesca Malvezzi and Serena Grassi for support in virtual docking analysis and, finally, to Michele Ibrahim Hamoude for support in expression and purification of the protein constructs. TG acknowledges computing time contributions by volunteers of the GPUGRID project, research funding from Acellera Ltd, and CINECA awards under the ISCRA initiative for the availability of high performance computing resources and support.

# Conflicts of interest

The authors declare no conflicts of interest.

# Tables

**Table 1. $K_d$ values of NF023 and of TRAF2 *vs* cIAP2-BIR1 forms measured in MST.** $K_d$ values were measured following the thermophoresis of the fluorescently labelled protein targets (wild type and mutant cIAP2-BIR1 forms) upon the addition of NF023 and TRAF2 at increasing concentrations. To assess NF023 ability to impair cIAP2-BIR1/TRAF2 binding, a 30' pre-incubation of the protein targets with NF023 at a final concentration of 500 µM was done prior to mixing with TRAF2 and measuring (with the exception of the S34A mutant, which did not bind NF023 nor TRAF2). Values and associated errors are calculated on triplicates.

| cIAP2-BIR1 forms (Targets) | $K_d$ (µM) vs | | |
|---|---|---|---|
| | NF023 | TRAF2 (no NF023) | TRAF2 after incubation with 500 µM NF023 |
| *Wild type* | 31 ± 9 | 39 ± 17 | No binding |
| *C28S* | 450 ± 160 | 12 ± 5 | 15 ± 8 |
| *Y31A* | 52 ± 19 | No binding | No binding |
| *S34A* | > 10000 | No binding | - |
| *Y56A* | > 10000 | 53 ± 23 | 40 ± 6 |

# Figures

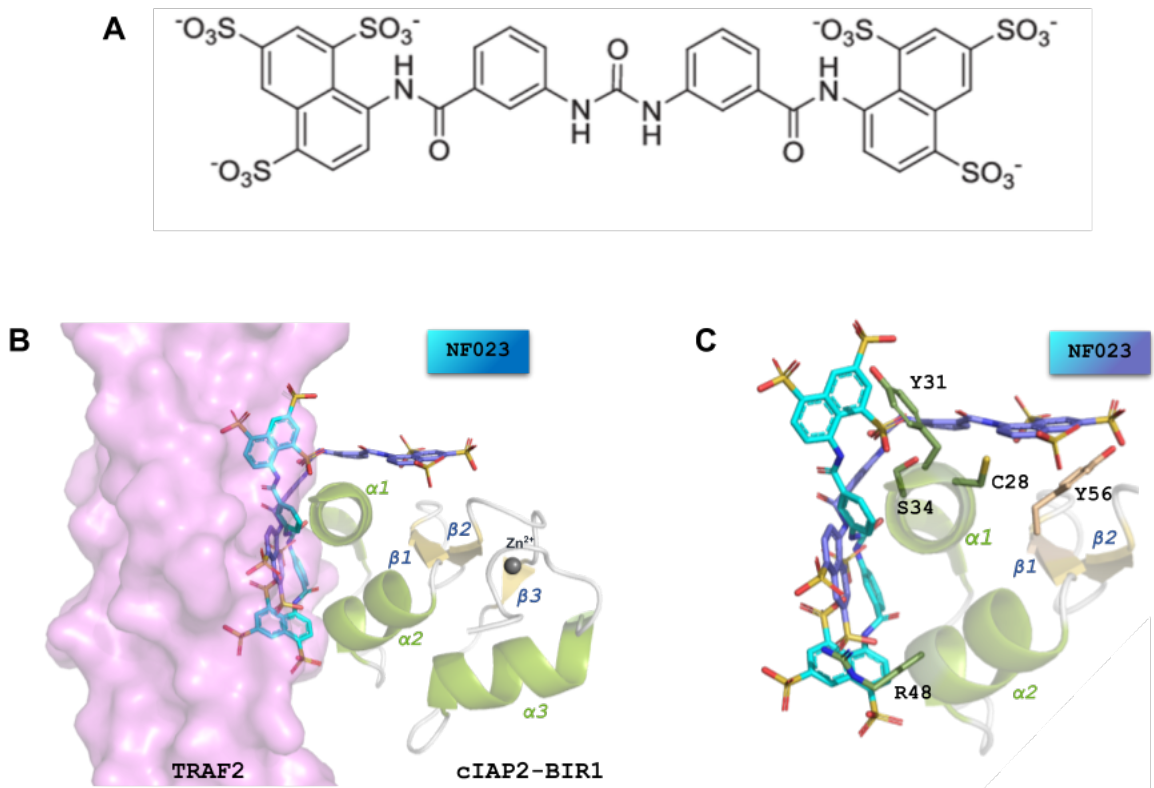

**Figure 1. NF023 and its identification as a disruptor of the cIAP2-BIR1/TRAF2 assembly.** A) Chemical structure of compound NF023. B) Two poses of NF023 (in cyan sticks, running along TRAF2 axis, and in light blue sticks, slightly bending over α1) result from a preliminary virtual screening, both covering the region of cIAP2-BIR1 (in cartoons; green: α-helices; yellow: β-strands) interacting with TRAF2 (pink surface). Reference structure PDB ID: 3M0A [7]. C) Zoomed view of the poses. Residues included in the mutagenesis experiments (C28, Y31, S34, R48, Y56) are highlighted in sticks. Figures were prepared with PyMOL (The PyMOL Molecular Graphics System, Version 2.0 Schrödinger, LLC).

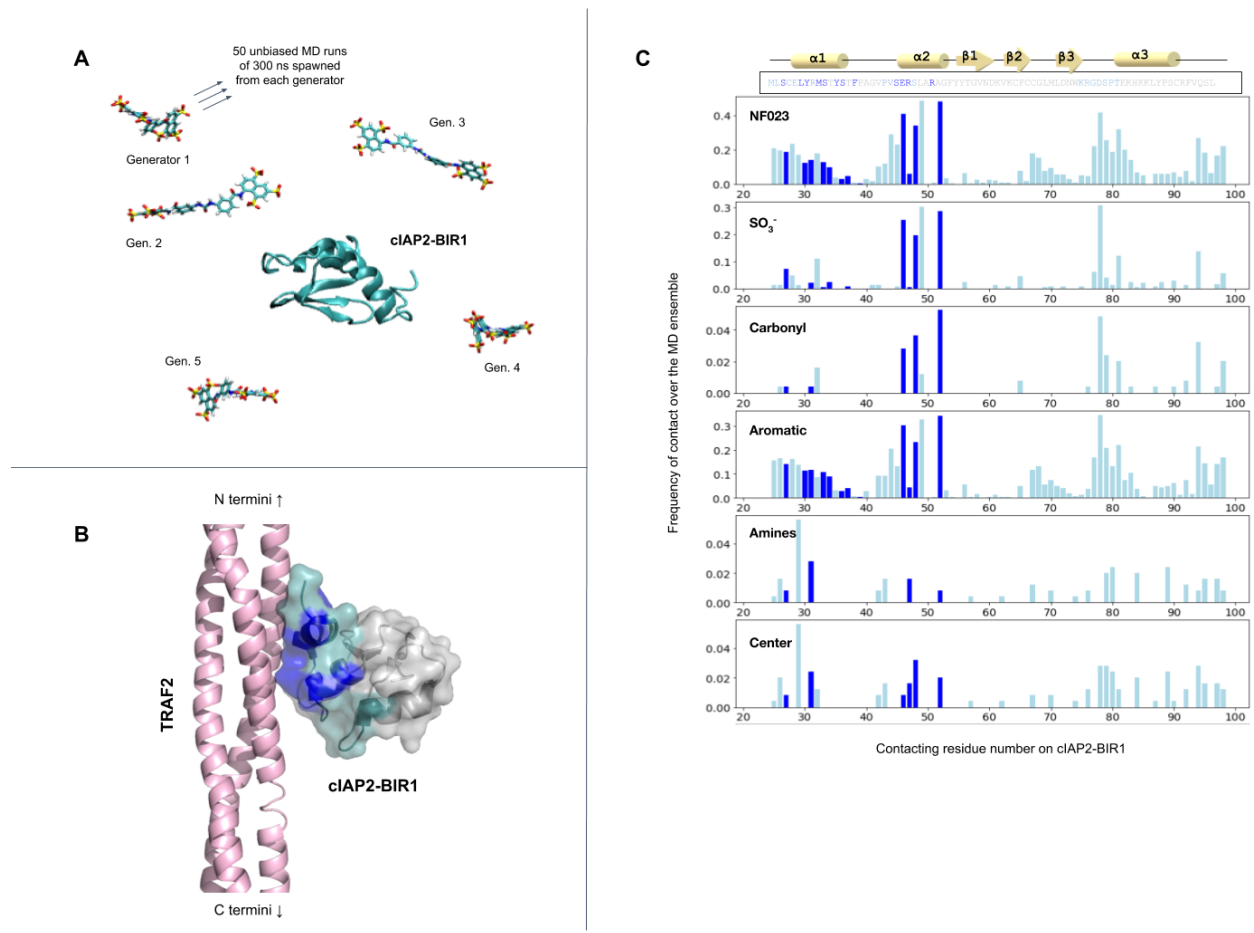

**Figure 2. Moiety-specific characterization of NF023:cIAP2-BIR1 binding from the ensemble of free-diffusion binding trajectories. A)** Illustration of the five initial configurations used for MD runs, or *generators*. Each generator differs for the location of NF023 in the solvent, chosen by randomly picking a point on a 34 Å-radius sphere centered on the protein. Each generator spawned 50 unbiased trajectories (250 total replicas) of 300 ns.. **B)** Three-dimensional view of cIAP2-BIR1 regions (blue surfaces) whose Cα atoms are less than four Å far from TRAF2 trace (pink cartoons), thus representing the core of the target PPI. The blue surface together with the light cyan region of cIAP2-BIR1 establish the most frequent NF023 contacts during dynamics simulations. **C)** The frequency of contacts between each of the indicated chemical moieties of NF023 and residues of the protein was estimated at the end of

248 simulations of $t$ = 300 ns. Residues at the TRAF2 interface are shown in dark blue, corresponding to blue surfaces in panel B. Overall, the interactions most frequently observed are hydrophobic at regions M25-A53 ($α$1 and $α$2) and K77-K85 (loop and $α$3), and polar amphipathic at S46-E47-R48-R52.

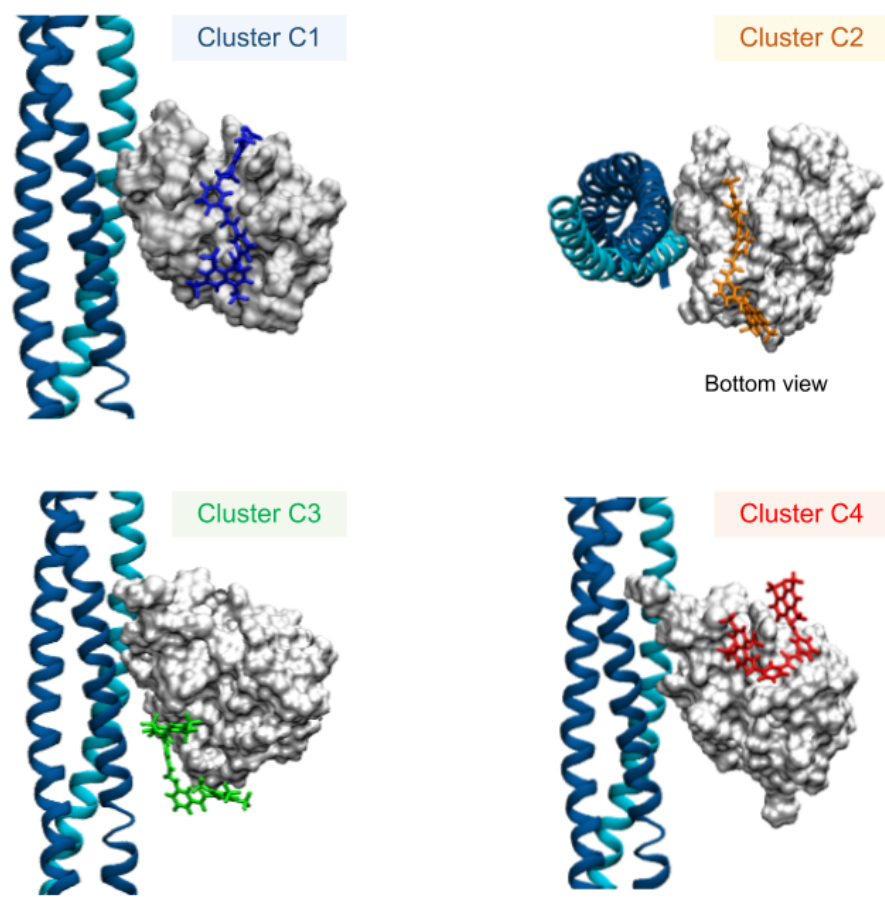

**Figure 3**. **Clusters of NF023 conformations across the final frames of the 248 independent replicas.** The 248 final frames, one per independent MD free-binding replica of 300 ns, were clustered with a cut-off threshold of 2 Å RMSD, yielding four poses (C1 to C4). Clusters 2 and 3 are in the "T" region, hence likely sterical disruptors of the TRAF2:cIAP2-BIR1 PPI assembly.

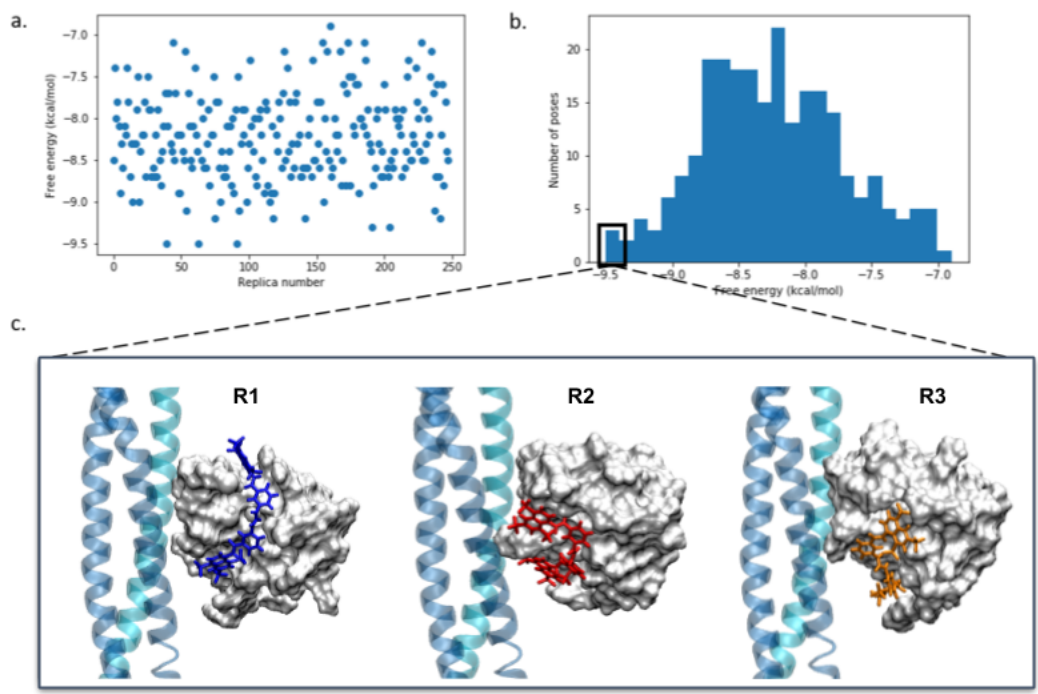

**Figure 4**. **NF023 "ensemble redocking" analysis with BIR1 conformations from 248 MD replicas.** (a) For each of the 248 replicas (horizontal axis), the last frame was extracted and Autodock Vina was used to dock NF023; the three best free energy values are reported. (b) Distribution of ΔG values found by the ensemble docking procedure. (c) The three overall best-scoring poses, with the free energy of -9.5 kcal/mol (black box) are represented together with the TRAF2:cIAP2-BIR1 complex.

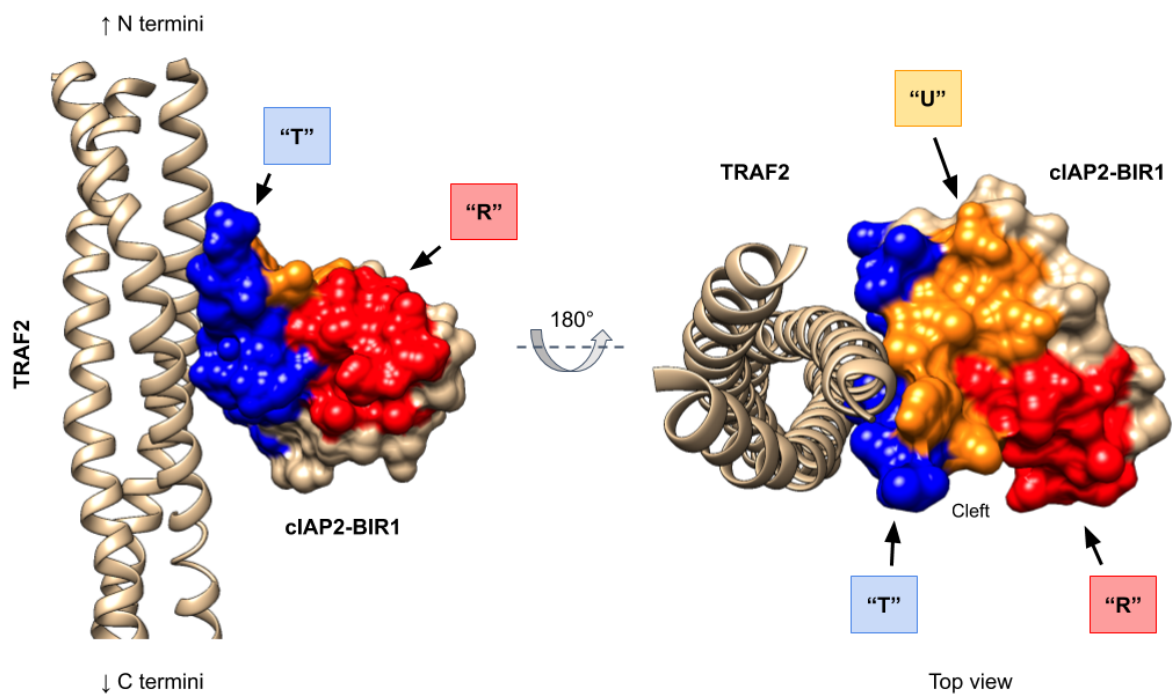

**Figure 5**. **The main interaction regions (*T, R, U*) identified on the cIAP2-BIR1 surface**. Surface regions "R" and "U" were identified on the basis of the final frames of the free-diffusion unbiased MD simulations, clustering and redocking analysis; interface "T" is the PPI interface region. The right-hand figure ("upper" region) is viewed from the N-terminal ends of TRAF2's coiled coils. See Table S2 for the residue/region assignments.

# Associated content

Supporting information available: description of the preliminary docking screen and of suramin analogs (supplementary texts S1 and S2); supplementary figures S1 to S7; supplementary tables S1 to S3. Supplementary data: PDB files (explicit solvent removed) of poses from the 4 clusters identifying over-represented final configurations in the MD ensemble; the three highest-energy poses from the ensemble redocking strategy; two poses identified in the preliminary screen.